\documentclass{aastex631}

\begin{document}

\title{Period-Luminosity Relationship for $\delta$ Scuti Stars Revisited}

\author[0000-0002-0196-9732]{Atila Poro}
\altaffiliation{poroatila@gmail.com}
\affiliation{Astronomy Department of the Raderon AI Lab., BC., Burnaby, Canada}

\author[0000-0002-9490-2093]{S. Javad Jafarzadeh}
\affiliation{Department of Physics, University of Texas at Dallas, 800 W. Campbell Rd., Richardson, TX 75080, USA}

\author{Roghaye Harzandjadidi}
\affiliation{Department of Theoretical Physics and Astrophysics, University of Zanjan, Zanjan, Iran}

\author{Mohammad Madani}
\affiliation{Independent researcher, Tehran, Iran}

\author{Elnaz Bozorgzadeh}
\affiliation{Department of Electronics and Telecommunications, Politecnico Di Torino, Turin, Italy}

\author{Esfandiar Jahangiri}
\affiliation{Independent researcher, Isfahan, Iran}

\author{Ahmad Sarostad}
\affiliation{Yazd Desert Night Sky Astronomy Institute, Yazd, Iran}

\author{Ailar Alizadehsabegh}
\affiliation{Department of Photonics, Tabriz University, Tabriz, Iran}

\author{Maryam Hadizadeh}
\affiliation{Khayyam Astronomy Association, Fars, Fasa, Iran}

\author{Mohammad EsmaeiliVakilabadi}
\affiliation{Khayyam Astronomy Association, Fars, Fasa, Iran}

\begin{abstract}
The Gaia DR3 parallax approach was used to estimate the absolute parameters of 2375 $\delta$ Scuti stars from the ASAS catalog. The selected stars have a variety of observational characteristics, with a higher than 80\% probability of being $\delta$ Scuti stars. We have displayed all the stars in the Hertzsprung-Russell (H-R) diagram along with the $\delta$ Scuti instability strip, the Zero Age Main Sequence (ZAMS), and the Terminal-Age Main Sequence (TAMS). Then, we determined which fundamental and overtone modes each star belongs to using pulsation constant ($Q$) calculations. In addition, we evaluated the parameters in the $Q$ calculation equation using three machine learning methods, which showed that surface gravity and temperature have the greatest effect on its calculation. The Period-Luminosity ($P-L$) relationship of the $\delta$ Scuti stars was also revisited. Eventually, using least squares linear regression, we made four linear fits for fundamental and overtone modes and updated their relationships.
\end{abstract}


%%%%%%%%%%%%%%%%% BODY OF PAPER %%%%%%%%%%%%%%%%%%
\vspace{1.5cm}
\section{Introduction}
The $\delta$ Scuti stars are sort of variable stars with a pulsation period range of 0.02 to 0.25 days that are mainly located in the lower part of the Cepheid instability strip along with SX Phe variables, which are mostly in the main sequence of star life. Their spectral type ranges from A to F (\citealt{2000ASPC..210..373M}). The two main groups of $\delta$ Scuti stars are the High Amplitude $\delta$ Scuti Stars (HADS) and the Low Amplitude Delta Scuti Stars (LADS). HADS are distinguished by their amplitude variations in brightness higher than 0.3 magnitude (\citealt{2017NewA...54...86J}); and LADS are known for having brightness amplitude variations less than 0.1 magnitude in $V$-band (\citealt{1996AA...307..539R}).

One of the fundamental properties of pulsating variable stars is the Period-Luminous relationship, as it was mentioned for the first time by \cite{1912HarCi.173....1L}. The first $P-L$ relation for $\delta$ Scuti stars was determined due to the stars' compliance with the same $P-L$ relation as Cepheids. In order to improve the $P-L$ relationship and be used as standard candles, \cite{1992AJ....103.1647F} examined combining a few Cepheids and $\delta$ Scuti stars.
In the following decades, other studies made an effort to improve the $P-L$ relationship for $\delta$ Scuti stars, such as \cite{laney2002dwarf}, \cite{mcnamara2004variable}, \cite{2007AJ....133.2752M}, \cite{2011AJ....142..110M}, \cite{2019MNRAS.486.4348Z}, \cite{2020MNRAS.493.4186J}, \cite{2021PASP..133h4201P}, \cite{2022MNRAS.516.2080B}, and \cite{2022ApJ...940L..25M}.

In this work, we aim to update the four relations for fundamental and three overtone modes with a large sample whose absolute parameters were estimated through one method. This paper is structured as follows: Section 2 introduces the dataset used in this study; Section 3 describes how to calculate the absolute parameters of the stars; Section 4 discusses pulsation modes and pulsation constants; Section 5 focuses on the update of the $P-L$ relationships; and Section 6 is the conclusion.

%%%%%%%%%%%%%%%%%%%%%%%%%%%%%%%%%%%%%%%%%%%%%%%%%%
\vspace{1.5cm}
\section{Dataset}
The $\delta$ Scuti stars, which were used as samples in this study, are from the ASAS catalog. These stars are classified as $\delta$ Scuti stars in the ASAS catalog, with a probability exceeding 80\%. Consequently, a total of 2375 $\delta$ Scuti stars form the basis of this sample. The periods of these stars in the ASAS catalog are between 0.031 and 0.199 days, while their apparent magnitudes span from 10.14 to 15.57.

Considering that Gaia's parallax was used in this study to estimate the absolute parameters, we checked the Re-normalised Unit Weight Error (RUWE) from Gaia DR3\footnote{\url{https://gea.esac.esa.int/archive/}} for these stars. All of the stars were in the appropriate range and below 1.4 (\citealt{lindegren2018re}).

%%%%%%%%%%%%%%%%%%%%%%%%%%%%%%%%%%%%%%%%%%%%%%%%%%
\vspace{1.5cm}
\section{Absolute Parameters Estimation}
The Gaia parallax method was used to estimate the absolute parameters of these stars (\citealt{2021PASP..133h4201P}). First, we calculated the absolute magnitudes of the stars using the apparent magnitude ($V$) from the ASAS catalog, the distance in parsec units ($d$) from Gaia DR3, and the extinction coefficient ($A_V$) using the dust-maps Python package of \cite{2019ApJ...887...93G}. Then, the bolometric correction (BC) from \cite{1996ApJ...469..355F}'s study, as corrected by \cite{2010AJ....140.1158T}, was applied to derive the bolometric absolute magnitude ($M_{bol}$) for each star. Using Pogson's relation (\cite{1856MNRAS..17...12P}), we computed the luminosity ($L$). The temperature ($T$) reported by TESS for these stars allowed us to calculate the radius ($R$) using $T$ and $L$. Additionally, we determined the mass ($M$) based on the relationship presented in \cite{cox2015allen}. We estimated the surface gravity on a logarithmic scale $log(g)$ using $M$ and $R$ parameters, which were in good agreement with the surface gravity provided by TESS for each star. The relations used in the order mentioned in the text, are as follows:

\begin{equation}\label{eq1}
M_V=V-5log(d)+5-(A_V)
\end{equation}

\begin{equation}\label{eq2}
M_{bol}=M_V+BC
\end{equation}

\begin{equation}\label{eq3}
\frac{L}{L_{\odot}}=10^{((M_{bol\odot}-M_{bol})/2.5)}
\end{equation}

\begin{equation}\label{eq4}
\frac{R}{R_{\odot}}=\sqrt{\frac{L/{L_{\odot}}}{(T_{eff}/T_{eff\odot})^4}}
\end{equation}

\begin{equation}\label{eq5}
logM=0.46-0.1M_{bol}
\end{equation}

\begin{equation}\label{eq6}
g=G_{\odot}(M/R^2)
\end{equation}

\begin{figure*}
\begin{center}
\includegraphics[width=\textwidth]{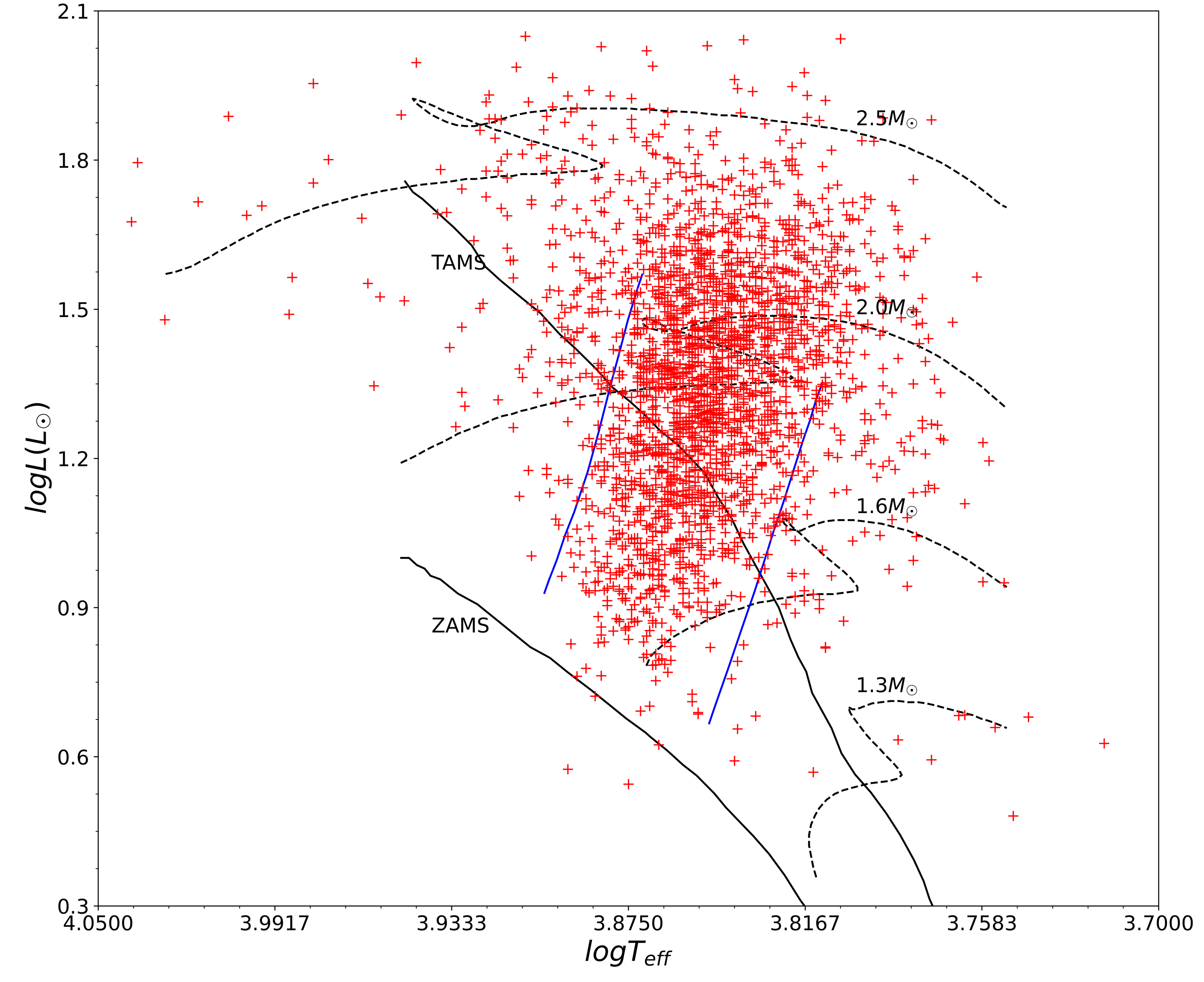}
    \caption{Location of 2375 stars on H-R diagram. A Monte Carlo process was used to evaluate the parameters' average uncertainty.}
\label{Fig1}
\end{center}
\end{figure*}

We placed all sample stars on the H-R diagram based on the estimated luminosity in this study and TESS temperature (Figure \ref{Fig1}). The H-R diagram displayed the instability strip of the $\delta$ Scuti stars (\citealt{2005AA...435..927D}, \citealt{murphy2019gaia}), the evolutionary paths of the stars (\citealt{2016MNRAS.458.2307K}), as well as the ZAMS and TAMS. As shown in Figure \ref{Fig1}, the stars are distributed across most of the $\delta$ Scuti stars' region.

%%%%%%%%%%%%%%%%%%%%%%%%%%%%%%%%%%%%%%%%%%%%%%%%%%
\vspace{1.5cm}
\section{Pulsation modes}
There are two categories of $\delta$ Scuti stars: radial and non-radial. The pulsation mode is classified as radial if the star maintains its spherical shape during pulsation; otherwise, the pulsing is non-radial. Additionally, pulsation modes are distinguished in terms of the quantum numbers ($n$, $l$, and $m$) that identify the star's geometry, where $n$ is the radial order, $l$ is the angular degree and $m$ is the azimuthal (\citealt{2017ampm.book.....B}, \citealt{2021PASP..133h4201P}).

The modes pressure ($p$) and gravity ($g$) describe the two different pulsation types identified in the $\delta$ Scuti stars. In the $p$-mode, pressure acts as a restoring force by sound waves with vertical movements for a star that has lost its equilibrium condition, and in the $g$-mode, gravity functions as a restoring force with horizontal movements for the star (\citealt{2015JApA...36...33J}). Identifying the $p$-mode order is necessary for studying the interior structure of $\delta$ Scuti stars. The pulsation constant is a crucial variable in this process.

The theoretical predictions for $Q$ were first made by \cite{1972ApJ...171...93C} and \cite{fitch19811}. Comparing measured periods to theoretically expected $Q$ values allows us to pinpoint certain radial and non-radial pulsation modes. These periods relate to the time a sound wave takes to travel from the star's center to its surface, expressed as $P\frac{1}{\sqrt{G\rho}}$, known as the Dynamic time scale (\citealt{aerts2010asteroseismology}). The pulsation constant $Q$, is defined as:

\begin{equation}\label{eq7}
Q=P\sqrt{\frac{\bar{\rho}}{{\bar{\rho}_{\odot}}}}
\end{equation}

where $P$ is period of pulsation in days, and $\bar{\rho}$ is the mean stellar density. Therefore, the pulsation modes in $\delta$ Scuti stars correspond to specific $Q$ values. According to \cite{1990DSSN....2...13B} and \cite{2021PASP..133h4201P}, $Q$ can be calculated by:

\begin{equation}\label{eq8}
log(\frac{Q}{P})=0.1(M_{bol}-M_{bol\odot})+0.5log(\frac{g}{g_{\odot}})+log(\frac{T_{eff}}{T_{eff\odot}})
\end{equation}

The calculation of the $Q$ value greatly depends on the stellar parameters. Due to this dependence, $Q$ values may have a fractional uncertainty of up to 18\% (\citealt{1990DSSN....2...13B}). The internal structure and evolution of $\delta$ Scuti stars can be studied using the $Q$ value, although doing so requires precise measurements and careful interpretation.

The $Q$ value for the sample stars was determined using equation 8. Then, we made an effort to evaluate equation 8's four parameters ($P$, $M_{bol}$, $T$, and $log(g)$) in terms of how they would impact the determination of $Q$. It should be noted that we used the value of temperature from TESS, the period from ASAS, and the values of bolometric absolute magnitude and surface gravity from the estimations of this study.

\begin{figure*}
\begin{center}
\includegraphics[width=\textwidth]{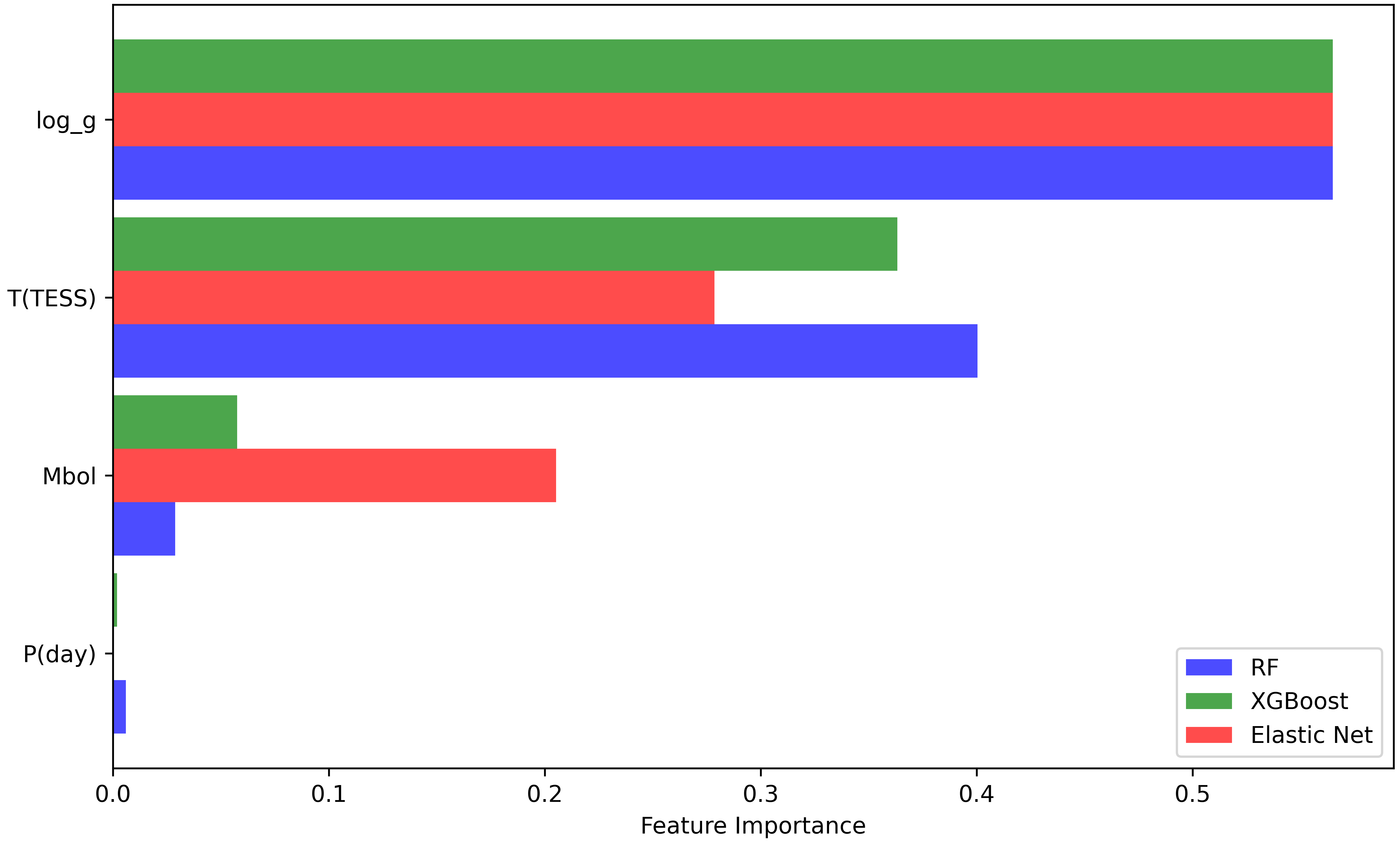}
    \caption{Histogram of measuring the importance of four parameters in the $Q$ relation using three machine learning methods.}
\label{Fig2}
\end{center}
\end{figure*}

We leveraged various methods, including Random Forest, XGBoost, and Elastic Net, to determine the relative importance of features in a machine learning problem. Before applying these techniques, we performed feature scaling on both the dependent variable, denoted as $Q$, and four independent variables using the StandardScaler. For each model, we conducted a thorough grid search Cross-Validation (CV) to identify the optimal hyperparameters that yield the highest accuracy. Subsequently, we fitted each model with these selected hyperparameters. Finally, we visualized the feature importance rankings. Since different models may provide results on varying scales, and our primary concern lies in the relative importance of these features, we standardized the value of the most crucial feature across all models. This standardization allows for a more straightforward comparison. Analyzing the results in Figure \ref{Fig2}, we observe a consistent feature ranking across all models. Notably, $log(g)$, $T$, $M_{bol}$, and $P$ are consistently listed in descending order of importance across the models.

%%%%%%%%%%%%%%%%%%%%%%%%%%%%%%%%%%%%%%%%%%%%%%%%%%
\vspace{1.5cm}
\section{Updated $P-L$ Relationship}
\begin{figure}
\begin{center}
\includegraphics[width=\textwidth]{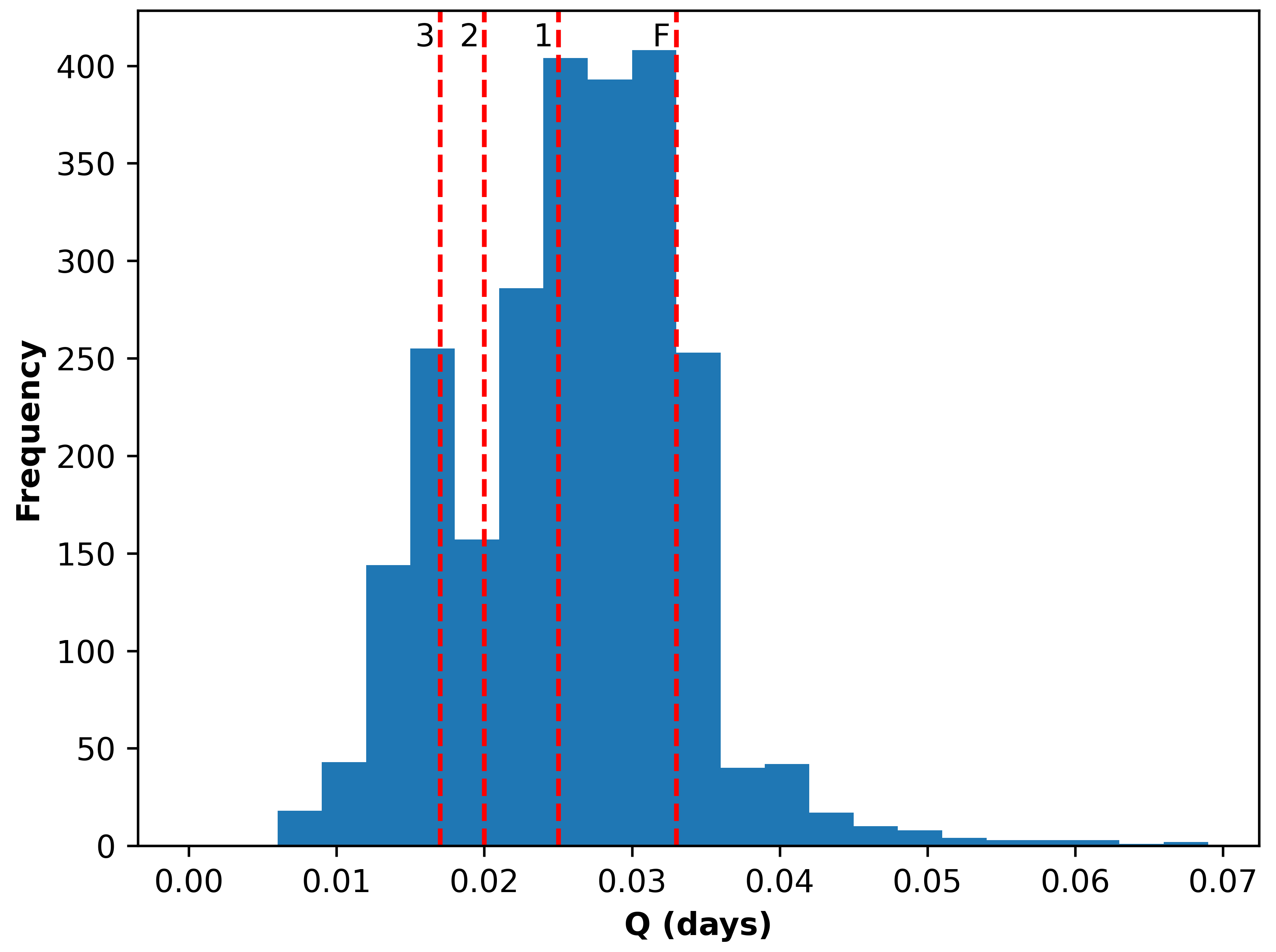}
    \caption{Histogram of the calculated $Q$s. Dashed lines show fundamental and three overtones, which are consistent with the North et al. (1997).}
\label{Fig3}
\end{center}
\end{figure}

To find the $P-L$ relationship for each of the three overtones and the fundamental, it is essential to categorize each star into its respective group. So, we utilized the histogram of the distribution of our calculated $Q$s for all stars, as illustrated in Figure \ref{Fig3}. In this histogram, dashed lines represent the fundamental and three overtones, as detailed in Figure \ref{Fig3} in \cite{1997ESASP.402..367N}. Except for the second overtone, which corresponds to a local minimum, the remaining three (two overtones and the fundamental) correspond to local maxima. The specific range of each group is outlined below:

\begin{equation}\label{eq9}
\left\{\begin{array}{l}
Fundamental: 0.027 \leq Q\\
First\ overtone: 0.021 \leq Q < 0.027\\
Second\ overtone: 0.018 \leq Q < 0.021\\
Third\ overtone: Q < 0.018
\end{array}\right.
\end{equation}

\begin{figure*}
\begin{center}
\includegraphics[width=\textwidth]{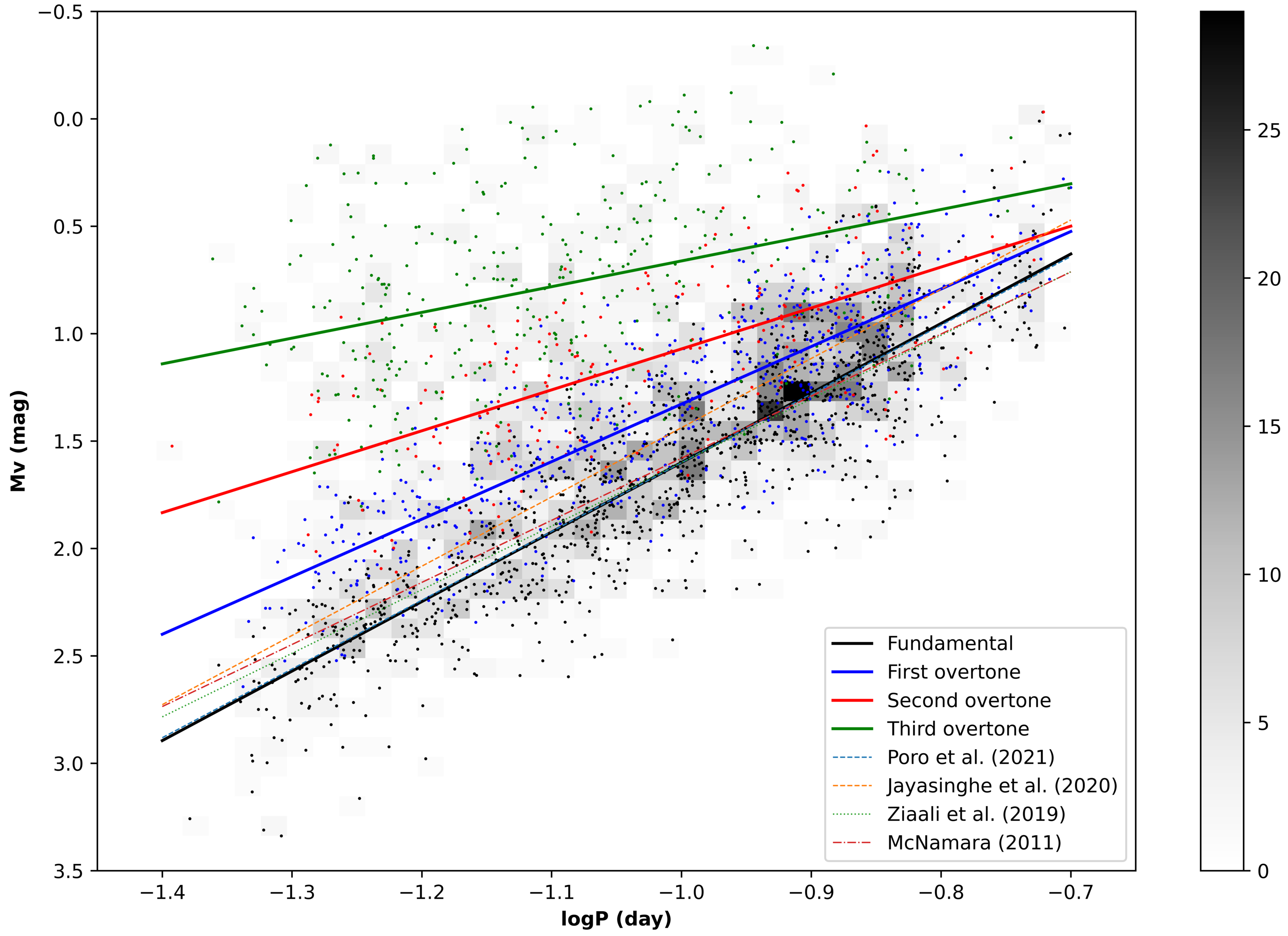}
\caption{2D histogram of ($M_V-logP$) and fitted lines of ($P-L$) relations for the studies $\delta$ Scuti stars of this paper. The fitted lines of four previous studies for the fundamental mode are included for reference.}
\label{Fig4}
\end{center}
\end{figure*}

After finding the regions of $Q$, each star can be visualized in a two-dimensional (2D) histogram based on its $M_V$ and $logP$ as in Figure \ref{Fig4}. In this histogram, fundamental, first, second, and third overtone stars are represented by black, blue, red, and green dots, respectively. Their population is visualized using a gray color map, with 35 bins and a peak near $M_V=1.25$ and $logP=-0.9$.
We used least squares linear regression to fit the best line to each of these four groups represented by the same color as the data points in Figure \ref{Fig4}. Moreover, the fitted $P-L$ relations from previous studies for the fundamental are presented as blue dashed for \cite{2021PASP..133h4201P}, orange dashed for \cite{2020MNRAS.493.4186J}, green dotted for \cite{2019MNRAS.486.4348Z}, red dashed-dotted for \cite{2011AJ....142..110M}. The updated $P-L$ relations for all four modes are as follows:

Fundamental:
\begin{equation}\label{eq13}
M_v=(-3.236\pm0.051)\times logP-(1.636\pm0.052)
\end{equation}

First overtone:
\begin{equation}\label{eq12}
M_v=(-2.679\pm0.069)\times logP-(1.351\pm0.070)
\end{equation}

Second overtone:
\begin{equation}\label{eq11}
M_v=(-1.905\pm0.152)\times logP(-0.834\pm0.157)
\end{equation}

Third overtone:
\begin{equation}\label{eq10}
M_v=(-1.198\pm0.157)\times logp-(0.536\pm0.174)
\end{equation}

\begin{figure*}
\begin{center}
\includegraphics[width=\textwidth]{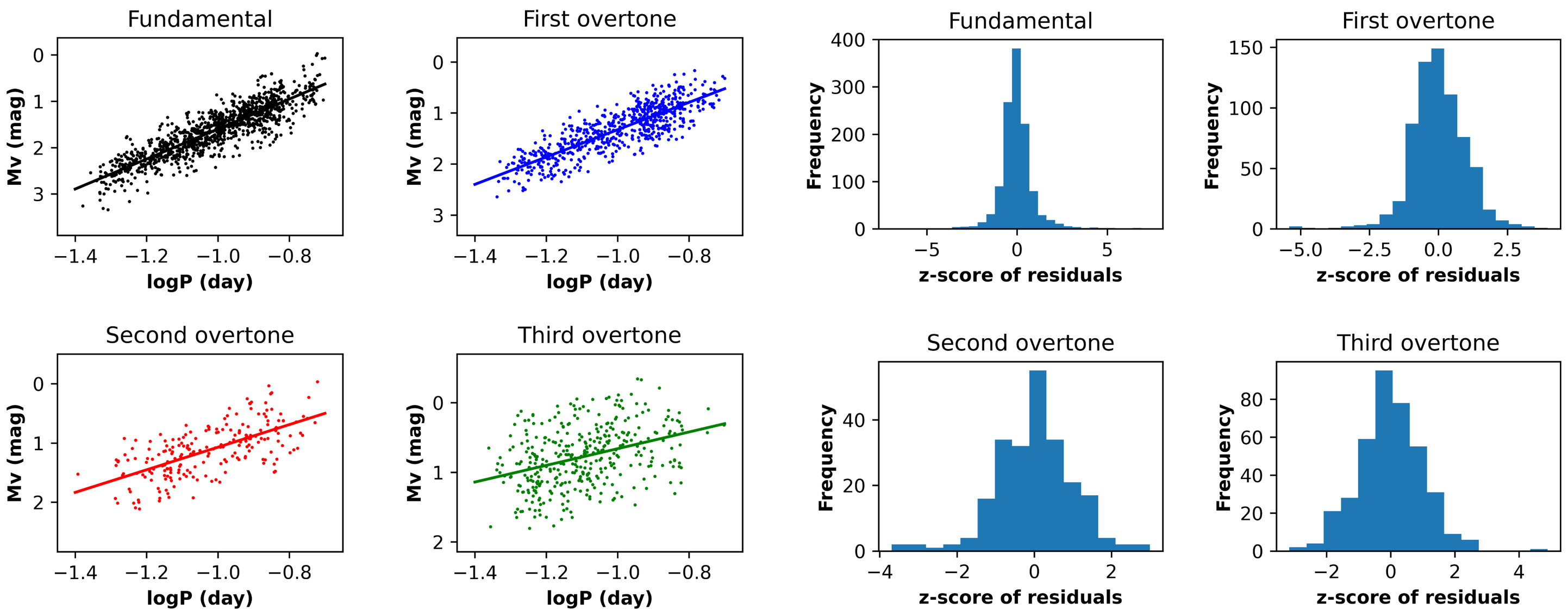}
\caption{Left: Separate linear fit on the fundamental and three overtones, Right: Four histograms of residuals of fitted equations 10–13.}
\label{Fig5}
\end{center}
\end{figure*}

Finally, we proceeded to compute the residuals by comparing our data points with the corresponding fitted equations. To enhance the interpretability of these residuals, we standardized them, setting their mean to zero and their standard deviation to one. The result for three overtones and fundamental is shown in Figure \ref{Fig5}.

%%%%%%%%%%%%%%%%%%%%%%%%%%%%%%%%%%%%%%%%%%%%%%%%%%
\vspace{1.5cm}
\section{Conclusion}
Using Gaia DR3's parallax, some catalogs, and databases, we calculated the physical parameters of 2375 $\delta$ Scuti stars. Stars with a high probability of being this type of pulsating star were selected. Given the period interval of less than 0.2 days for the stars, this is the largest sample for $\delta$ Scuti stars. The results obtained for the $log(g)$ parameter show good agreement with the TESS findings.

The positions of the stars in the H-R diagram are displayed in Figure \ref{Fig1}. A large number of stars are between the two theoretical lines of the $\delta$ Scuti star range. It is worth mentioning that \cite{2011A&A...534A.125U} revealed that some $\delta$ Scuti stars could be found outside their instability strip. However, it should be considered that most of the stars outside the $\delta$ Scuti instability strip might be hybrid objects.

We then examined the pulsation modes of the stars. Also, we employed different machine learning methods to assess the impact of the parameters involved in the $Q$ constant. This investigation revealed that $Q$ was least affected by the pulsation period in all three machine learning methods, which is probably due to the short interval of period changes in most $\delta$ Scuti stars.

We revised the $P-L$ relationship in this work. To do this, we estimated $Q$, which differed marginally from the \cite{1997ESASP.402..367N} study (Figure \ref{Fig3}).
Then, we separated the stars for the first, second, and third overtones as well as the fundamental mode. For them, a linear fit was the best choice. So, four updated relationships with the sample of this study were presented (Equations 10-13).

Comparing the results to \cite{2021PASP..133h4201P}, we realized that, while this study encompasses a significantly larger sample, our findings were consistent with theirs, updating the relations to greater accuracy.
Although the updated fundamental equation remains almost the same as in \cite{2021PASP..133h4201P}, we notice a slightly greater change in overtones. The reason for that could be that in \cite{2021PASP..133h4201P}'s study, according to the total number of 520 stars, their number in overtones was much less and more scattered than in fundamental.

Comparing these results to \cite{2021PASP..133h4201P}, we observe that, despite this study encompassing a significantly larger dataset of $\delta$ Scuti stars, our findings align consistently with theirs, updating the relations to greater accuracy.
While the updated fundamental equation remains almost the same as in \cite{2021PASP..133h4201P}, we notice a slightly more pronounced change in overtones. This difference may stem from the fact that, in \cite{2021PASP..133h4201P}'s study, where a total of 520 stars were examined, the number of stars in overtones was considerably fewer and more scattered compared to the fundamental mode. This disparity could have influenced the precision of their results.

We also utilized machine learning methods to assess the suitability of the non-linear fit (\citealt{2024RAA....24a5002P}). The evaluation, based on Root Mean Square Error (RMSE) values, indicated that the linear fit emerged as the best-fitting model.
This is a significant issue because a recent study by \cite{2022ApJ...940L..25M} discovered nonlinear behavior for about 3700 $\delta$ Scuti stars from extra-galactic systems within the $P-L$ relationship. However, According to the 2375 stars examined in this study, the Period-Luminosity relationship for Milky Way $\delta$ Scuti stars can still be adequately described by a linear fit.

%%%%%%%%%%%%%%%%%%%%%%%%%%%%%%%%%%%%%%%%%%%%%%%%%%
\begin{acknowledgements}
This investigation has made use of data from the European Space Agency mission Gaia (\url{http://www.cosmos.esa.int/gaia}), processed by the Gaia Data Processing and Analysis Consortium (DPAC, \url{http://www.cosmos.esa.int/web/gaia/dpac/consortium}).
We express our gratitude to Dr. Filiz Kahraman Aliçavuş for providing scientific and effective comments.
\end{acknowledgements}

%%%%%%%%%%%%%%%%%%%%%%%%%%%%%%%%%%%%%%%%%%%%%%%%%%
\vspace{1.5cm}
\section*{Availability}
All estimates of the physical parameters of the 2375 $\delta$ Scuti stars are included in the online version of this study.

%%%%%%%%%%%%%%%%%%%%%%%%%%%%%%%%%%%%%%%%%%%%%%%%%%
\vspace{1.5cm}
\section*{ORCID iDs}
\noindent Atila Poro: 0000-0002-0196-9732\\
Seyed Javad Jafarzadeh: 0000-0002-9490-2093\\
Roghaye Harzandjadidi: 0000-0002-1836-0958\\
Mohammad Madani: 0000-0003-4705-923X\\
Elnaz Bozorgzadeh: 0009-0009-7143-263X\\
Esfandiar Jahangiri: 0000-0002-1576-798X\\
Ahmad Sarostad: 0000-0001-6485-8696\\
Ailar Alizadehsabegh: 0000-0001-5768-0340\\
Maryam Hadizadeh: 0000-0003-1493-0295\\
Mohammad EsmaeiliVakilabadi: 0009-0004-9852-8002\\

%%%%%%%%%%%%%%%%%%%%%%%%%%%%%%%%%%%%%%%%%%%%%%%%%%
\vspace{1.5cm}
\bibliography{Ref}{}
\bibliographystyle{aasjournal}

%%%%%%%%%%%%%%%%%%%%%%%%%%%%%%%%%%%%%%%%%%%%%%%%%%

\end{document}